\documentclass[useAMS,usenatbib]{mn2e}
\usepackage{amsmath,graphicx,psfig}

\def\arcmin{^{\prime}}

\def\micron{$\rm{\umu}$m}
\def\microns{$\rm{\umu}$m}

\title[Extending PLE models into the mid-IR, far-IR and sub-mm]{Extending PLE models into the mid-IR, far-IR \& sub-mm} \author[M. D. Hill \& T. Shanks] {Michael D. Hill$^{1}$\thanks{e-mail:
m.d.hill@durham.ac.uk} \& Tom Shanks$^{1}$ \\ $^{1}$Dept.\ of Physics, Science Laboratories, University of Durham, Durham, DH1 3LE, UK}

\begin{document}

\pagerange{\pageref{firstpage}--\pageref{lastpage}} \pubyear{2010}

\maketitle

\label{firstpage}

\begin{abstract}Simple pure luminosity evolution (PLE) models, in which galaxies
brighten at high redshift due to increased star-formation rates (SFRs), are
known to provide a good fit to the colours and number counts of galaxies
throughout the optical and near-infrared. We show that optically defined
PLE models, where dust reradiates absorbed optical light into infrared spectra
composed of local galaxy templates, fit galaxy counts
and colours out to 8\microns\ and to at least $z\approx2.5$. At
24--70\microns, the model is able to reproduce the observed source counts with
reasonable success if 16\% of spiral galaxies show an
excess in mid-IR flux due to a warmer dust component and a higher SFR, in line
with observations of local starburst galaxies. There remains an under-prediction
of the number of faint-flux, high-$z$ sources at 24\microns, so we explore how
the evolution may be altered to correct this. At 160\microns\ and longer
wavelengths, the model fails, with our model of normal galaxies accounting for
only a few percent of sources in these bands. However, we show that a PLE model
of obscured AGN, which we have previously shown to give a good fit to
observations at 850\microns, also provides a reasonable fit to the \emph{Herschel}/BLAST 
number counts and redshift distributions at 250--500\microns. In the context of a 
$\Lambda$CDM cosmology, an AGN contribution at 250--870\microns\  would remove the need to
invoke a top-heavy IMF for high-redshift starburst galaxies, although the
excellent fit of the galaxy PLE model at shorter wavelengths would still
need to be explained.

\end{abstract}

\begin{keywords} infrared: galaxies -- dust, extinction -- galaxies: evolution -- submillimetre\end{keywords}

\section{Introduction}

The process by which galaxies formed and evolved into the systems we see today remains one of the most important questions in modern astronomy. We do not currently have a comprehensive picture of how and when most of the stellar content of the universe was assembled. 

In studies of galaxy evolution there are, broadly, two ways to proceed. One is the \emph{ab initio} approach, in which galaxy evolution is modelled using a chosen prescription for processes including the collapse and merging of dark matter halos, gas heating and cooling, star-formation and feedback (e.g.\ Cole et al., 2000). The other approach is more observationally led, beginning with a description of the local luminosity functions (LFs) and spectral energy distributions (SEDs) of galaxies and `winding the clock back' to build up a picture of the high-redshift universe (Bruzual \& Kron, 1980; Koo, 1981; Shanks et al., 1984; Pozzetti et al., 1998). These approaches are often called `forward evolution' and `backward evolution' respectively (for a review, see \S5.2 of Hauser \& Dwek, 2001). While forward evolution models can potentially offer greater insight into the physical processes that may be driving galaxy evolution, the advantage of backward evolution models is that they begin with a phenomenological description of the local universe, which is comparatively well understood and well constrained.

The semi-analytical models used to simulate structure formation in the $\Lambda$CDM cosmology employ a forward evolution, \emph{ab initio} approach. In these models, galaxy formation is a hierarchical process, in which stellar mass is built up over extremely long timescales through the merging of smaller structures (Cole et al., 2000). However, hierarchical models have faced significant observational challenges. To summarise a wide range of studies, the broad observational picture is that the most massive galaxies host old stellar populations (Franx et al., 2003; Cimatti et al., 2004; Glazebrook et al., 2004; Thomas et al., 2005; Bernardi et al., 2006) apparently indicating a high redshift of formation, and show little evolution in their number density since $z\approx1$ (Wake et al., 2006; Brown et al., 2007; Cool et al., 2008), suggesting no significant dynamical evolution in the last several Gyr. These observations are contrary to the expectation from hierarchical evolution, where massive structures would be expected to have assembled since $z\approx1$. 

The observations are more in line with the predictions of passive evolution models, which assume that galaxies formed early and were in place at some high redshift, thereafter evolving to the present day only in terms of a decreasing star-formation rate. Instead of a hierarchical, merger-driven process of structure formation, passive evolution models assume galaxies formed from massive gas clouds in comparatively short-lived `monolithic collapse' events.

In this paper we use a model which employs the backward evolution approach and pure luminosity evolution, beginning with a local LF and evolving it to higher redshift using $k$- and/or $(k+e)$-corrections with no density evolution. We model galaxy evolution by simply assuming that star-formation declines at different rates in early- and late-type galaxies. This model has been used extensively in the past at optical and near-infrared (NIR) wavelengths (Metcalfe et al., 1995, 2001, 2006; McCracken et al., 2000) where, with modifications (see Metcalfe et al., 2006), it has been successful in matching observations from the $U$ to the $K$ band and out to $z\approx4$. We now test the model's predictions against observations in the mid-IR, far-IR and sub-mm.

These long wavelength regimes are important to galaxy evolution studies. Observations with the \emph{COBE} satellite in the 1990s confirmed the existence of a cosmic infrared background (CIB) which peaks in the far-IR at $\lambda\approx100$--300\microns\ (Puget et al., 1998; Fixsen et al., 1998; Hauser et al., 1998). It is now clear that the CIB, which arises from the re-emission by dust of absorbed radiation, has a total intensity roughly equal to that of the extragalactic background at optical wavelengths. Since infrared sources contribute only $\approx 30$\% of the total energy output of the local universe (Soifer \& Neugebauer, 1991), there is expected to be significant evolution in the populations which contribute to the infrared background. It is clear, then, that an understanding of the evolution of galaxies at long infrared wavelengths is important to studies of the star-formation and, since AGN may contribute to this background, accretion history of the universe.

Other authors have presented models which give a good fit to mid-IR, far-IR and sub-mm observations, using both hierarchical forward evolution models (e.g.\ Granato et al., 2000; Lacey et al., 2008) and phenomenological backward evolution models (e.g.\ Chary \& Elbaz, 2001; Rowan-Robinson, 2001, 2009; Lagache, Dole \& Puget, 2003; Lagache et al., 2004).

Typically, these models are designed specifically to match infrared observations. The hierarchical $\Lambda$CDM model requires a top-heavy IMF to fit the FIR/sub-mm data (Baugh et al., 2005; Lacey et al., 2008), and the phenomenological models tend to employ a range of different spectral types, including luminous starbursts, or allow for strong evolution in the luminosity functions. Here our approach differs: all aspects of our model, including the modest amount of dust absorption, were defined in the optical bands. The surprise has been that this simple model continued to represent the data well as it was applied at high redshifts and to longer, near-IR wavelengths. We now extend this much further, out to the mid-IR, far-IR and sub-mm regimes, following Busswell \& Shanks (2001). Determining where and how the model breaks down is instructive, since this may indicate a requirement for dynamical evolution or for another population to dominate the counts, e.g. AGN (see Hill \& Shanks, 2010).

This paper is structured as follows. In \S\ref{sec:data} we discuss the datasets used; in \S\ref{sec:models} we discuss our galaxy models. In \S\ref{sec:colours} we investigate galaxy colours. \S\ref{sec:countsmir} and \S\ref{sec:countsfir} deal with predictions of our PLE models for number counts at mid-IR and far-IR/sub-mm wavelengths respectively. In \S\ref{sec:disc} and \S\ref{sec:summ} we present a discussion and a summary of our findings.

Magnitudes throughout the paper are quoted in the Vega system unless otherwise stated. For consistency with the work of Metcalfe et al.\ we use a cosmology with $H_0 = 50$ km s$^{-1}$ Mpc$^{-1}$, $\rm{\Omega}_m = 0.3$ and $\rm{\Omega}_{\rm{\Lambda}} = 0.7$. 



\begin{figure}
\includegraphics[width=80.mm]{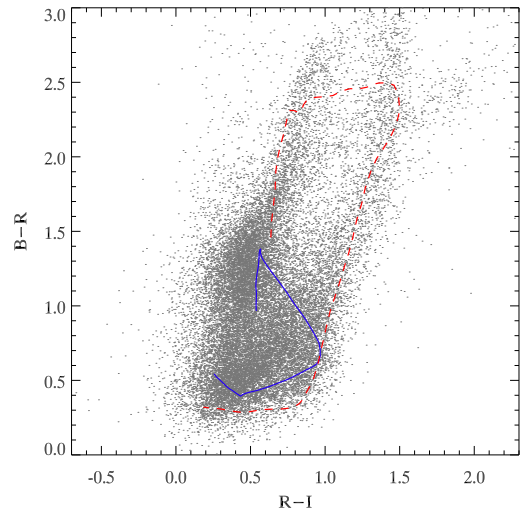}
\caption{A BRI colour-colour diagram showing the predicted redshift tracks for the $\tau=9$ spirals (solid blue line) and the $\tau=1$ ellipticals (dashed red line). The data shown are $R<25$ galaxies from  deep Subaru imaging of the GOODS-N field. Our models match well the looping features which are clearly visible in the data.}
\label{fig:bri}
\end{figure}

\section{Data} \label{sec:data}

To test our model predictions of colours at \emph{Spitzer} IRAC wavelengths, we make use of data from the $\approx200$ arcmin$^2$ GOODS-North survey. This field is well suited since it has some of the deepest NIR and IRAC imaging available over a relatively wide area and has also been targetted in several spectroscopic follow-up surveys. For photometry we use the catalogue recently presented by Wang et al.\ (2010), who combine public \emph{Spitzer} IRAC data (Dickinson et al., in prep.) with a new ultradeep $K$-band image, using the $K$ positions as priors and detecting robust IRAC counterparts. We get redshifts for these sources by matching to redshift catalogues from Cohen et al.\ (2000), Cowie et al.\ (2004) and Wirth et al.\ (2004). We supplement this with the catalogue presented by Reddy et al.\  (2006), which includes spectroscopic redshifts and $K$-band and IRAC fluxes for 388 high-$z$ galaxies in GOODS-N. Our total redshift sample includes 1,617 galaxies.

Numbers counts of galaxies in the \emph{Spitzer} IRAC (3.6, 4.5, 5.8 and 8.0\micron) and MIPS (24, 70 and 160\microns) bands have been measured across a wide flux range by surveys which combine observations from several fields of differing width and depth in order to achieve good sampling of both the faint and bright end of the source counts. At IRAC wavelengths, we use the published counts of Fazio et al.\ (2004), who provide counts from three regions ranging from an ultradeep $5\arcmin\times10\arcmin$ field to the wide $3^{\circ}\times3^{\circ}$ Bo\"otes field. A similar selection of fields was used by Papovich et al.\ (2004) and Dole et al.\ (2004) in the MIPS bands, with the combination of deep and wide fields allowing them to determine number counts across 3 decades of flux. A more recent survey by Bethermin et al.\ (2010) took this approach further, combining MIPS observations across 9 fields of various depths covering $\approx54$ deg$^2$ in total. We use all of these datasets, as well as our own counts from public GOODS-N imaging in the IRAC bands (Dickinson et al., in prep.) and at 24\micron\ (Chary et al., in prep.) using SExtractor (Bertin \& Arnouts, 1996) on the image frames.

We also compare our model to observations at 60\microns, where counts are available from wide-field IRAS surveys (Lonsdale et al., 1990; Rowan-Robinson et al., 1991; Gregorich et al., 1995), although these observations do not reach the same depth achieved by the more recent \emph{Spitzer} surveys.

Finally, we use recent galaxy surveys in the sub-mm bands at 250, 350 and 500\microns. These are the wavebands probed by the SPIRE detector on \emph{Herschel} (Pilbratt et al., 2010) and by \emph{Herschel's} balloon-borne forerunner, BLAST (Patanchon et al., 2009). We make use of BLAST number counts (Patanchon et al., 2009) and redshift distributions (Chapin et al., 2010; see also Dunlop et al., 2009) from observations covering $\approx10$ deg$^2$. We use SPIRE number counts from the HerMES survey, currently covering $\approx20$ deg$^2$, both as published by Oliver et al.\ (2010) and also as presented by Dunlop et al.\ (2010).


\section{Model} \label{sec:models}

Here we give a brief summary of the galaxy evolution model we employ, and we review some previous results from this model at optical/NIR wavelengths. For a more detailed description of the model we refer the reader to Metcalfe et al.\ (2006) and references therein.

\subsection{Galaxy types}

Our basic galaxy model assumes only two populations, broadly corresponding to early-types (old, red, passive systems) and late-types (young, blue, star-forming spirals). Both types are modelled with exponentially decreasing star formation rates (SFRs), i.e.\ $\rm{SFR}(t) \propto \rm{e}^{-t/\tau}$. The early- and late-type galaxies are distinguished by different characteristic e-folding times, $\tau$: for early-types we use $\tau=1$ Gyr, for late-types $\tau=9$ Gyr. Based on this formula we use Bruzual \& Charlot (2003; hereafter BC03) models to produce template spectral energy distributions (SEDs) for these two galaxy types, assuming a Chabrier IMF. 

Assuming a bimodal galaxy population is clearly a simplification, but it is not without merit. Optical colour-magnitude and colour-colour diagrams show evidence of strong bimodality, with a so-called `red sequence' and `blue cloud', the former consisting primarily of bulge-dominated galaxies with old stellar populations and the latter of young, disk-dominated, star-forming galaxies (Strateva	et al., 2001). Such plots can often be seen to show two curving tracks, and these are well matched by models using a $\tau=1$ and a $\tau=9$ population (Metcalfe et al., 2001); this is clear from Fig.\ \ref{fig:bri}.

\subsection{Dust spectrum}

In order to apply this model at infrared wavelengths, a prescription for dust absorption and re-radiation is required. We invoke dust only for the late-type galaxies. Following Metcalfe et al.\ (2001) and Busswell \& Shanks (2001), we apply a 1/$\lambda$ extinction law normalised at 4500\AA\ in the B band with some normalisation $A_B$. That is, the flux at a given wavelength after absorption, $F_{\lambda}$, relates to the unabsorbed flux $F_{\lambda,0}$ according to equation \ref{eq:flambdaabs}.

\begin{equation}
F_{\lambda}=10^{-0.4 \times A_B \times \frac{4500\rm{\AA}}{\lambda} } F_{\lambda,0}
\label{eq:flambdaabs}
\end{equation}


\begin{figure}
\includegraphics[width=70.mm]{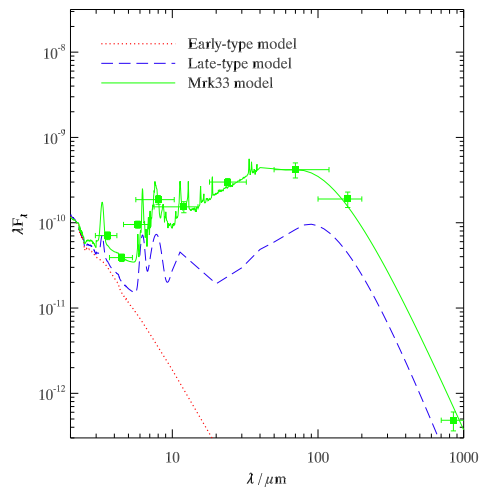}
\caption{Infrared SEDs for three galaxy templates. Our standard, bimodal models comprises the early-type (red dotted line) and late-type (blue dashed line) galaxies; the early-types are dustless so we see only the long-wavelength tail of the stellar emission, while the late-types include dust with a schematic representation of the PAH features (see text). Also shown is an SED based on the starburst galaxy Mrk33 (green solid line), which we introduce at $\lambda \ge 24$\microns\ in place of a fairly small fraction of the normal late-types. The PAH spectrum for Mrk33 comes from IRS observations and the green datapoints show broadband photometry for Mrk33. All models and data are normalised to be equal at 2\microns; the $y$-axis is in arbitrary units of $\lambda F_{\lambda}$.}
\label{fig:modelseds}
\end{figure}

The absorbed radiation is re-emitted by the dust in the infrared. The primary component of this emission is a modified Planck function, which has some given dust temperature and opacity. We apply an opacity law with the standard form $\kappa \propto \lambda^{-\beta}$, where a value of $\beta=0$ would correspond to a perfect blackbody. In our model we set $\beta=1.5$ (Dunne \& Eales, 2001).

Emission from polycyclic aromatic hydrocarbons (PAHs) is an important spectral feature between 3 and 13 microns (Leger \& Puget, 1984; Allamandola et al., 1985). In our spiral galaxy model we include a fairly schematic representation of the PAH features, designed to broadly resemble the spectra of galaxies observed in the Spitzer Infrared Nearby Galaxies Survey (SINGS;  Kennicutt et al., 2003), presented by Draine et al. (2007). We show the IR spectra of our early- and late-types in Fig.\ \ref{fig:modelseds}. While the modelling of the PAH features in the late-types is somewhat rudimentary, we find it is sufficient to reproduce the observed colours and sources counts of galaxies which sample this part of the spectrum, so we consider it sufficient for the purposes of this paper.

Fig.\ \ref{fig:modelseds} also includes a third SED, which represents a starburst component modelled on Markarian 33 (Mrk33). This is an additional spectral type which we introduce when dealing with observations at $\lambda \ge 24$\microns\ (see \S\ref{ssec:24}, where the motivation for including this component is discussed).

We assume a dust temperature of 30K, which is found to be a typical temperature of interstellar dust in star-forming galaxies (Farrah et al., 2003; Pope et al., 2006; Coppin et al., 2008; Elbaz et al., 2010). We take $A_B=0.3$ magnitudes for the normalisation of the $1/\lambda$ dust absorption law; this is the value determined by Metcalfe et al.\ (2001) when using this model at optical wavelengths, and is a fairly conservative amount of extinction. The total integrated flux of the dust emission, in the form of both the PAH features and the blackbody, is normalised to be equal to the total absorbed flux.


\begin{figure}
\includegraphics[width=80.mm]{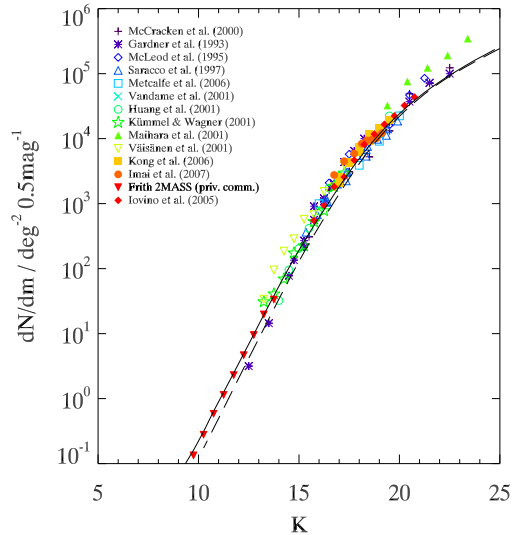}
\caption{Differential number counts in the K band. The solid line is the prediction of the model used in this work; the dashed line shows the model of Metcalfe et al.\ (2006) who use an $x=3$ IMF for early-types as described in the main text.}
\label{fig:kcnts}
\end{figure}


\begin{figure}
\includegraphics[width=80.mm]{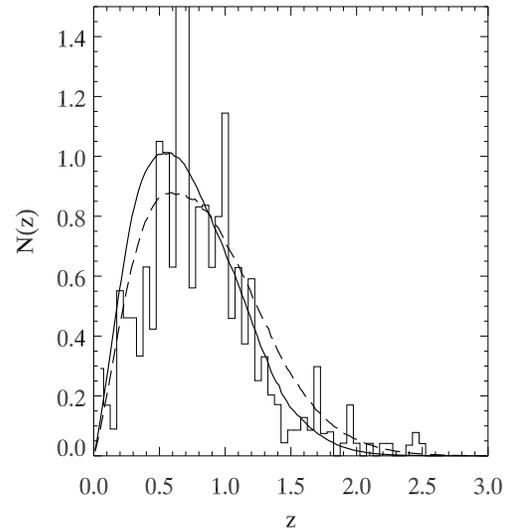}
\caption{The redshift distribution of sources in the K band. The histogram shows observational data from the K20 survey (Cimatti et al., 2002) down to a magnitude limit of $K=20$. The solid and dashed lines are as in Fig.\ \ref{fig:kcnts}}
\label{fig:k20}
\end{figure}

\begin{table}
\caption{Luminosity function parameters. These are the same as those used by Metcalfe et al.\ (2006). When shifting the LFs to another band, we use 2 colours, from a $\tau=1$ Gyr model for early-types (E/S0 and Sab) and $\tau=9$ Gyr model for late-types (Sbc, Scd and Sdm) to get new values fof $M^*$ for a given band, as shown in Table \ref{t-cols}.} 
\centering
\label{t-lf}
\begin{tabular}{|c|c|c|c|}
\hline
Type & $\Phi^* / Mpc^{-3}$ & $\alpha$ & $M^*_{K}$  \\
\hline
E/S0 & $9.27\times10^{-4}$ & $-0.7$ & $-24.92$ \\
Sab & $4.63\times10^{-4}$ & $-0.7$ & $-24.78$   \\
Sbc & $6.20\times10^{-4}$ & $-1.1$ & $-24.83$   \\
Scd & $2.73\times10^{-4}$ & $-1.5$ & $-24.34$   \\
Sdm & $1.36\times10^{-4}$ & $-1.5$ & $-23.71$ \\
\hline
\end{tabular}
\end{table}

\subsection{Initial mass function and luminosity function}

For each galaxy type, we use a luminosity function (LF) defined locally in the B
band and correct it into the appropriate IR band using a rest-frame $z=0$
colour. The resulting LF parameters for the $K$ band are given in Table \ref{t-lf}.
Then using $k+e$ corrections determined by the BC03 evolution code, the
LF is evolved back from the present day. The model is a simplification of the
model of Metcalfe et al.\ (2001, 2006) since rather than 5 independent colours, one for each
type, we use only two, one for all of the early-types (E/S0 and Sab) and one for all of 
the late-types (Sbc, Scd and Sdm). This is justified on the basis that the
band-band colour differences in the NIR are smaller than in the optical. 
The $z=0$  colours are taken from the BC03 model predictions. 
Table \ref{t-cols} shows the values of $M^*$ after shifting into
MIR/FIR/sub-mm bands. On this basis, we determine  predictions for the observed 
galaxy number counts and colours.

\begin{table*}
\caption{Characteristic magnitudes, $M^*$, for the luminosity function of our galaxy types in all of the mid-IR, far-IR and sub-mm wavebands used in this paper. Since the ellipticals are dustless in our model their contribution falls essentially to zero in the FIR/sub-mm, therefore these colours are omitted. As described later in the paper, at 24\microns\ and above we replace the Scd population with a population based on the starburst galaxy Mrk33. Note that for consistency with Metcalfe et al.\ we use Vega colours out to 24\microns. At 60\microns\ and above this is infeasible, so the magnitudes given for these bands are in the AB system.}
\label{t-cols}
\begin{tabular}{|c|c|c|c|c|c|c|c|c|c|c|c|c|c|}
\hline
Type & $M^*_{3.6}$ & $M^*_{4.5} $ & $M^*_{5.8}$ & $M^*_{8.0}$ & $M^*_{24}$ & $M^*_{60}$ & $M^*_{70}$ & $M^*_{160}$ & $M^*_{250}$ & $M^*_{350}$ & $M^*_{500}$ & $M^*_{850}$ \\
\hline
E/S0 &  $-25.08$ & $-25.10$ & $-25.17$ & $-25.27$ & $-25.36$ & -- & -- & -- & -- & -- & -- & -- \\
Sab  &  $-24.94$ & $-24.96$ & $-25.03$ & $-25.13$ & $-25.22$  & -- & -- & -- & -- & -- & -- & -- \\
Sbc  & $-25.29$ & $-25.27$ & $-26.49$ & $-27.81$ & $-30.19$ & $-26.07$ & $-26.41$ & $-26.61$ & $-25.61$ & $-24.68$ & $-23.67$ & $-22.56$ \\
Scd  & $-24.80$ & $-24.78$ & $-26.00$ & $-27.32$ & -- & -- & -- & -- & -- & -- & -- & -- \\
Starburst & -- & -- & -- & -- & $-32.92$ & $-27.58$ & $-27.74$ & $-27.60$ & $-26.69$ & $-25.87$ & $-25.00$ & $-23.44$ \\
Sdm  & $-24.17$ & $-24.15$ & $-25.37$ & $-26.69$ & $-29.07$ & $-24.95$ & $-25.29$ & $-25.49$ & $-24.49$ & $-23.56$ & $-22.55$ & $-21.44$ \\
\hline
\end{tabular}
\end{table*} 


\begin{figure*}
\includegraphics[width=130.mm]{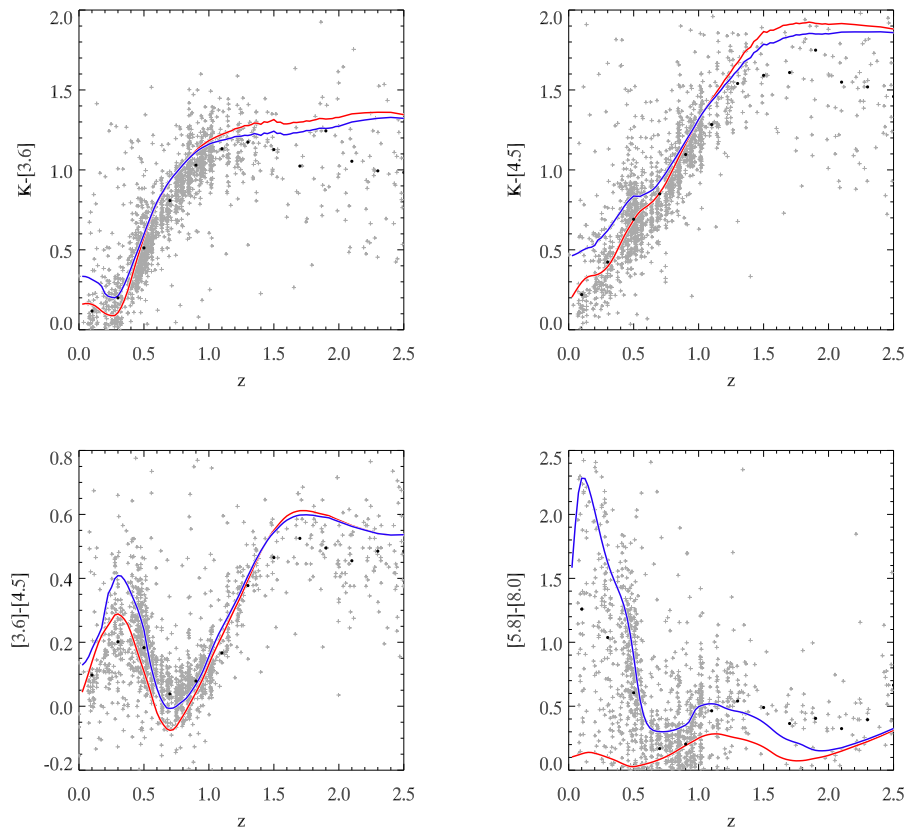}
\caption{Colour-redshift plots showing how our models compare to $K<22.5$ sources from GOODS-N spectroscopic surveys. Red and blue lines are PLE redshift tracks for our early-type and late-type models respectively. Black circles indicate the median of the data in $\Delta z=0.2$ bins. Overall there is a good agreement between the data and the model predictions.}
\label{fig:czplots}
\end{figure*}


\begin{figure}
\includegraphics[height=0.85\textheight]{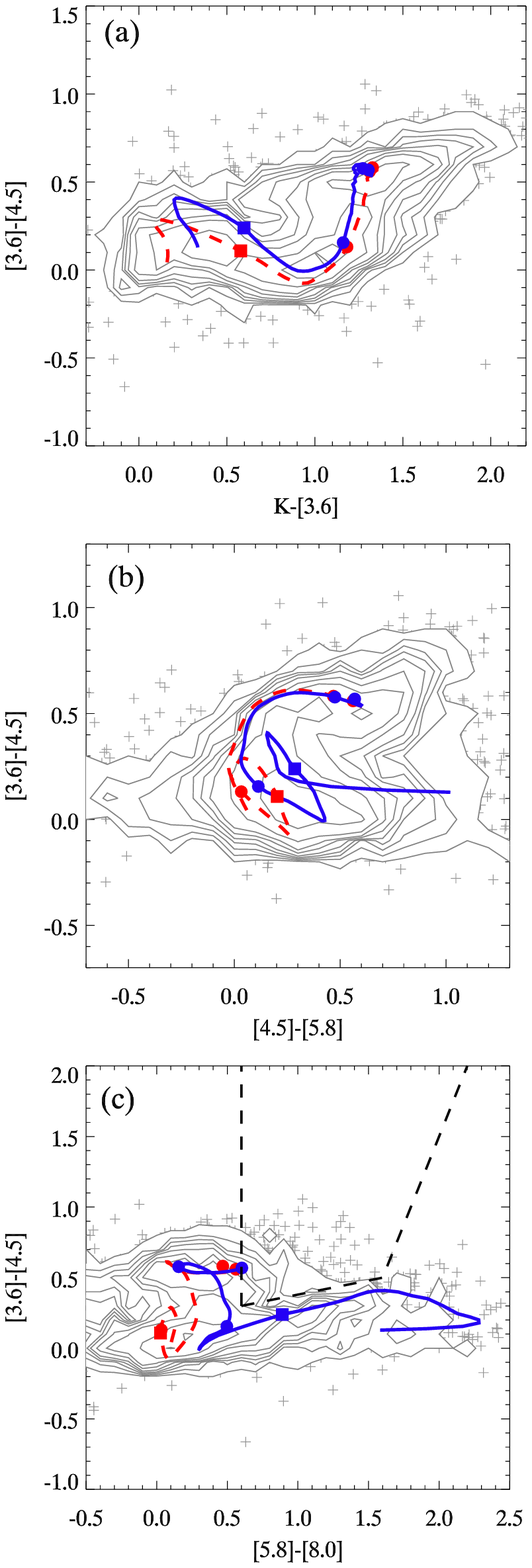}
\caption{Colour-colour plots showing how our models compare to observational data from GOODS-N. The contours show $K<21$ galaxies, with the lowest contour plotted where there are 2 objects in a $0.1\times0.1$ mag bin -- below this level galaxies are shown as individual crosses. The red (dashed) and blue (solid) lines are PLE redshift tracks for our early-type and late-type models respectively. The square marker on each track indicates $z=0.5$ and the circles further along the lines mark $z=1$, 2 and 3. For reference, in panel (c) we also show the selection criteria for the AGN wedge (dashed line; Stern et al., 2005).}
\label{fig:ccplots}
\end{figure}


Metcalfe et al.\ (2001, 2006) used a Salpeter IMF (Salpeter, 1955; $x=1.35$) for the $k+e$ corrections of spirals but 
used an $x=3$ IMF, cut at $M>0.5M_{\odot}$, for the $k+e$
corrections of early-types. This was to reduce the passive evolution in the
K-band to almost the level of the K-band $k$-correction to prevent too many
$z>1$ galaxies being predicted at $K<20$, which would be the case for the
Salpeter IMF. Here we are using the Chabrier IMF (Chabier, 2003; $x=1.3$ at 
$M>1M_{\odot}$) with an SFR e-folding time of $\tau=9$ Gyr for the spirals; this 
produces $k+e$ corrections very similar to the Salpeter with $\tau=9$ Gyr. 
For the early-type galaxies, no $x=3$ (or Scalo, 1986) IMF was easily 
available from BC03 and so here for predicting number counts and $n(z)$ in the K 
band and redward, we have simply assumed the $k$-correction
from the Chabrier IMF, with $\tau=1$ Gyr rather than $\tau=2.5$ Gyr 
to account for the degeneracy between IMF slope and SFR in the $k+e$ predictions 
(Metcalfe et al., 2001). We have checked that these are good  approximations to 
the $x=3$ predictions in the near-/mid-IR bands. Ultimately, however, the choice of
IMF and SFR for early-types has little impact on our results since the early-types are
taken to be dustless, and their contribution is therefore negligible in most of the 
mid-IR, far-IR and sub-mm wavebands which are the focus of this paper.

In Fig.\ \ref{fig:kcnts} we show the model predictions for the K band number counts generated
in this way and we note the model fits the counts very well across the full
range of magnitudes. We also see that the model agrees well with the $x=3$ model
of Metcalfe et al. In Fig.\ \ref{fig:k20} we show the K band redshift distribution compared
to the data from the K20 survey (Cimatti et al., 2002) where again there is
excellent agreement between the model and the data, with the fit being comparable to
that found with the $x=3$ model of Metcalfe et al.


\section{Galaxy colours} \label{sec:colours}

In Fig.\ \ref{fig:czplots} we present our model predictions for colour-redshift tracks in near-/mid-IR bands. The predictions are compared to observed colours of $K<22.5$ galaxies in the GOODS-N catalogue of Wang et al.\ (2010), described in \S\ref{sec:data}. Overall, the agreement between the data and our model in Fig.\ \ref{fig:czplots} is very good, suggesting that a simple optically-defined PLE galaxy model can match colours in the mid-IR out to at least $\lambda=8$\micron\ and $z=2.5$. The predicted spiral galaxy tracks in Fig.\ \ref{fig:czplots} show a `bump' in $K-[3.6]$ at $z\approx0.1$ and in $K-[4.5]$ at $z\approx0.5$. These arise from the PAH emission feature at $\lambda=3.3$\microns\ passing through the 3.6 and 4.5 micron bands, respectively. In the $K-[4.5]:z$ plot there is clear evidence of this bump in the data, well matched by the model, however the corresponding bump is not as clearly seen in the $K-[3.6]:z$ plot. This PAH line is also responsible for the significant peak seen in the $[3.6]-[4.5]$ colour at $z\approx0.4$, which again is replicated by the data.

The difference between the early- and late-type models is most pronounced in the $[5.8]-[8.0]$ colour (Fig.\ \ref{fig:czplots} lower-right panel), because the 5.8\micron\ and 8.0\micron\ bands sample the region of the spectrum where the contribution from dust is important relative to that from starlight. Since our early-types are modelled without dust it is the predicted colours of the late-types which give a better fit to most of the data here.

In Fig.\ \ref{fig:ccplots} we show colour-colour plots in the IRAC bands and in each case it is clear that the models are successful in matching the data, which again comes from Wang et al.\ (2010) and is cut to $K<21$. Figs.\ \ref{fig:czplots} and \ref{fig:ccplots} indicate that by taking simple empirical spectra and evolving them back to higher $z$ it is possible to get a reasonably accurate prediction for the average mid-IR colours of galaxies at a given redshift.  Using colours to select galaxies in a specified redshift range has become a powerful tool, for example in the selection of Lyman-break galaxies (LBGs; e.g. Steidel et al., 2003; Adelberger et al., 2004, Bielby et al., 2010), distant red galaxies (DRGs; Franx et al., 2003), extremely red objects (EROs; e.g. Roche et al., 2002), pBzK and sBzK galaxies (Daddi et al., 2004), and ``BM/BX'' galaxies (Steidel et al., 2004). Our PLE models may provide a simple underlying framework in which these colour selections can be better understood.


\begin{figure*}
\includegraphics[width=130.mm]{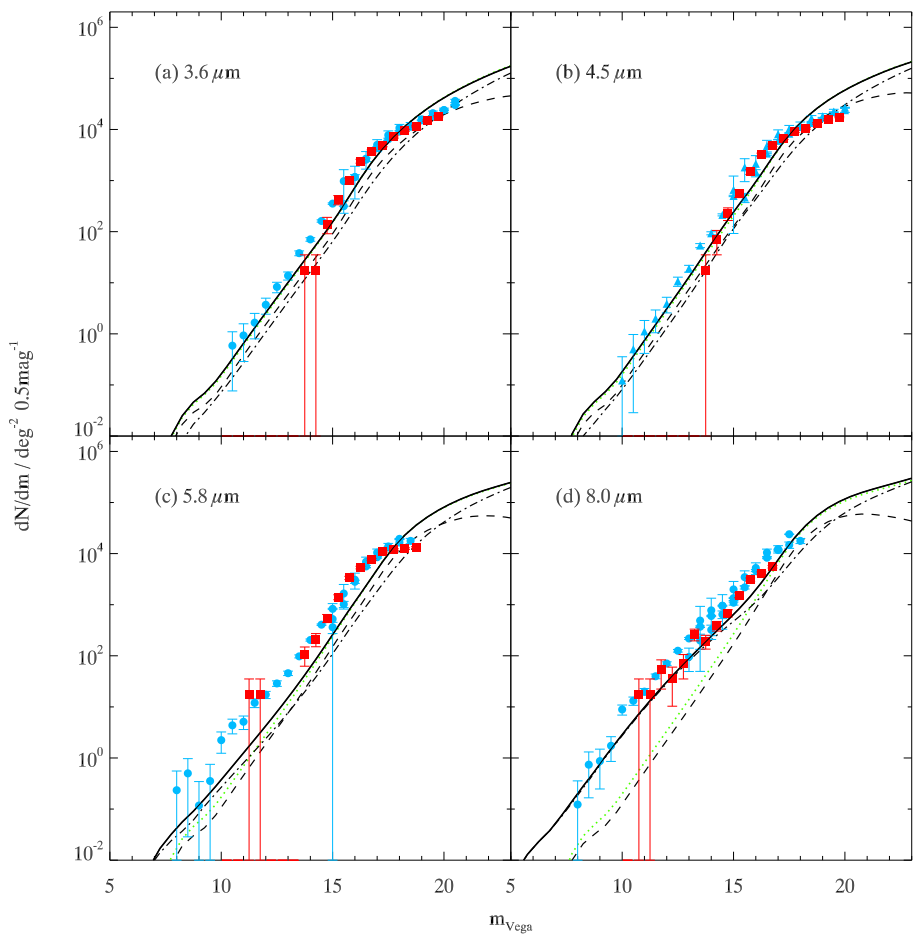}
\caption{Differential number counts in IRAC's 4 channels: (a) 3.6\microns, (b) 4.5\microns, (c) 5.8\microns\ and (d) 8.0\microns. The blue circles are data from Fazio et al.\ (2004) and the red squares are our own counts from public GOODS-N data. The solid black line shows the prediction of our model. The dot-dashed and dashed black lines indicate the contributions from late- and early-type galaxies, respectively. The green dotted line shows the combined contribution of these populations but from the stellar light only, i.e. without the emission from dust.}
\label{fig:iraccnts}
\end{figure*}


The $\rm{[3.6]}-\rm{[4.5]}:\rm{[5.8]}-\rm{[8.0]}$ colour-colour plot (Fig.\ \ref{fig:ccplots}\emph{c}) has been used widely to select populations at different redshifts (Stern et al., 2005; Yun et al., 2008; Dey et al., 2008; Devlin et al., 2009). The data in this plot visibly separate into two `tiers' with the upper tier, redder in [3.6]--[4.5], hosting higher-$z$ objects. Our models predict that the upper and lower tiers should divide galaxies above and below $z=1.3$; this is consistent with the finding of Devlin et al.\ (2009) that it cut galaxies at $z\approx1.2$, especially given that they found $\approx15$\% contamination of higher-$z$ sources in the lower tier and lower-$z$ sources in the upper tier. Stern et al.\ (2005) identified an `AGN wedge' in the upper tier, a region they suggest is populated by active galactic nuclei. For reference we have marked this region in Fig.\ \ref{fig:ccplots}\emph{c}; it is located redward of the high-$z$ end of our galaxy models in [5.8]--[8.0].

Essentially, simple colour-redshift tracks present a viable way for observers to select galaxies in a given $z$-range. We find that sources which lie on the upper ($z>1.3$) tier of the $\rm{[3.6]}-\rm{[4.5]}:\rm{[5.8]}-\rm{[8.0]}$ plot also lie around the $1.3<z<3$ track in, for example, a $B-R:R-I$ or $R-I:I-K$ plot, so we emphasise that colour selection techniques in the context of PLE models or otherwise can be used successfully and consistently all the way through the optical, near-IR and out to mid-IR wavelengths.

In summary, we have demonstrated that mid-IR colour-redshift data show no contradiction or inconsistency with the predictions of optically-defined galaxy evolution models with $\tau=1$ Gyr for early-types and $\tau=9$ Gyr for late-types.


\section{Mid-IR number counts} \label{sec:countsmir}

\subsection{IRAC bands} \label{ssec:irac}

Fig.\ \ref{fig:iraccnts} shows the number counts of galaxies at 3.6, 4.5, 5.8 and 8.0\microns. The predictions of our model are compared to data from Fazio et al.\ (2004), who combine observations from three regions ranging from an ultradeep $5\arcmin\times10\arcmin$ field to the wide $3^{\circ}\times3^{\circ}$ Bo\"otes field, in order to achieve good sampling of both the faint and bright end of the counts. We also compare to counts of GOODS-N sources by using SExtractor on the public \emph{Spitzer} imaging of the that field (Dickinson et al., in prep.), which agree well with Fazio et al.'s measurements. In both datasets stars have been excluded. 

Broadly speaking, the agreement between the model and the data is good. Although the model appears to overpredict the counts at the faint end in panels (a)--(c), the data here suffer from confusion. The fit at 5.8\microns\ is poorer than in the other bands, possibly due to our somewhat unsophisticated modelling of the PAH features, which may have a more noticeable effect at 5.8\microns\ than elsewhere. While a more refined model of these features may improve the prediction, we do not consider this adaptation to be necessary for the purposes of this paper. Indeed, we note that the fit to the 5.8\micron\ data is very similar to that achieved by the $\rm{\Lambda}$CDM hierarchical formation models presented by Lacey et al.\ (2008). 

\begin{figure}
\includegraphics[width=80.mm]{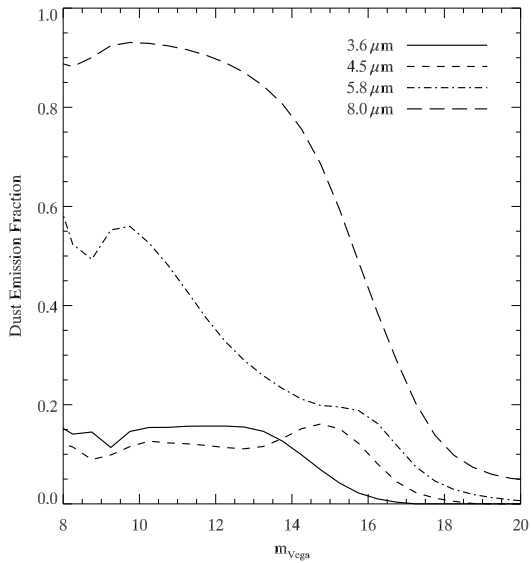}
\caption{The fraction of the galaxy number counts accounted for by dust emission, rather than stellar light, in each of the IRAC channels, based on the models in Fig.\ \ref{fig:iraccnts}. The shift from starlight-dominated to dust-dominated emission appears to occur around $\lambda\approx5$--6\microns.}
\label{fig:dustfrac}
\end{figure}

The plots also show the contribution made in each band by stellar light only (as opposed to dust). At 3.6\microns\ and 4.5\microns\ the stellar component accounts for the almost all of the counts, while in the longer wavelength bands it yields an underprediction. In Fig.\ \ref{fig:dustfrac} we show the fraction of the total predicted counts arising from dust emission in each of the IRAC bands, clearly demonstrating the transition from starlight-dominated to dust-dominated galaxy counts as we move from the near- to the mid-infrared.

In each band the very faintest source counts remain starlight-dominated, even at 8\microns. The model suggests that direct stellar emission accounts for only $\approx10$\% of the 8\micron\ number counts at magnitudes brighter than $m=13$ (a flux of $S_{8\umu\rm{m}}\ga0.4$ mJy), but makes up the dominant fraction ($>90$\%) of 8\micron\ sources fainter than $m=18$ ($S_{8\umu\rm{m}}\la4$ $\umu$Jy). 

Overall, we have shown that a basic PLE model, originally designed to match optical $B$-band counts, continues to provide a reasonably good description of the observed source counts out to the mid-IR wavelength regime at 8\microns\ if a modest amount of dust absorption and re-emission for spiral galaxies is included.


\subsection{24 microns} \label{ssec:24}

\begin{figure}
\includegraphics[width=80.mm]{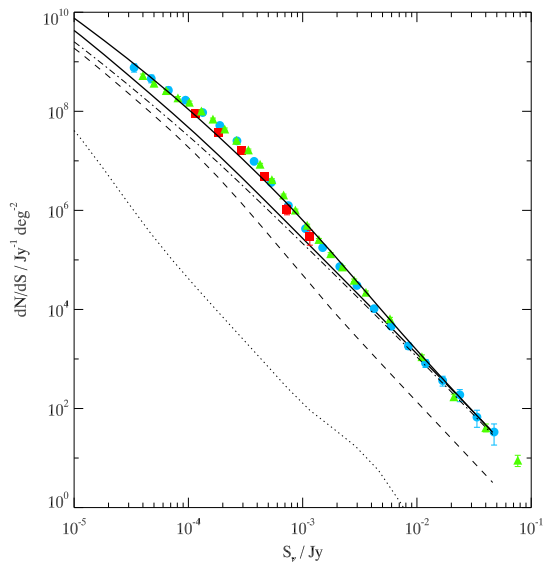}
\caption{Differential number counts at 24\microns. Blue circles are data from Papovich et al.\ (2004), green triangles are from B\'ethermin et al.\ (2010) and red squares are our own counts from public GOODS-N MIPS data. The lower solid line is the prediction of our model, with the `normal' spiral galaxy contribution shown as a dashed line, Mrk33 galaxies shown by a dot-dashed line and the negligible early-type contribution shown as a dotted line. Mrk33 galaxies clearly dominate the counts, with the normal galaxy component underpredicting the data by a factor of $\sim5$. The bright end is extremely well matched, but at the faint end the model is deficient by a factor of a few. We note that the spirals match the shape of the counts well, so we show also the prediction of the model where instead of Mrk33 we use an SED with the bright-end ($z=0$) normalisation of Mrk33, but the $k+e$ evolution of the normal spirals. This is the upper solid line.}
\label{fig:24cnts}
\end{figure}

In Fig.\ \ref{fig:24cnts} we show 24\micron\ number counts, with data from Papovich et al.\ (2004) and our own source counts from public GOODS-N 24\micron\ data (Chary et al., in prep.). We find that our model of normal galaxies, using the same dust parameters which had reasonable success at $3.6<\lambda<8.0$ \microns\, underpredicts the number counts by a factor of 4--5 (dashed line in Fig.\ \ref{fig:24cnts}), accounting for only 22\% of $F_{24}>30$ $\umu$Jy sources. This, then, is the wavelength regime where our basic model first breaks down.

However, we find that a relatively simple alteration to the model addresses this problem somewhat. We introduce a subset of our spiral galaxies, modelled on the dust features of Mrk33, as presented in the SINGS spectroscopic survey of local galaxies. This survey uses two-component dust models to match observed mid-IR SEDs of local galaxies (Draine et al., 2007). In most cases the dust emission is dominated by dust in the diffuse ISM, but some galaxies' dust emission is dominated by a warmer dust component, attributable primarily to photo-disassociative regions (PDRs), where the dust is exposed to a stronger intensity of radiation. These sources are found to have typically higher star-formation rates.

The PDR dust is hotter and thus its emission peaks at shorter wavelength (see Fig.\ \ref{fig:modelseds}). As a result of this strong PDR contribution, therefore, these galaxies exhibit a significant excess in flux in the mid-infrared. We find that they can make a significant contribution in the $\lambda \ge 24$\microns\ regime, without adversely affecting the fit of the model at shorter optical to mid-IR wavelengths.

18\% of the 65 SINGS galaxies discussed in Draine et al.\ (2007) are of this type; these include Mrk33, NGC2798, NGC3049 and Tol89. We modelled an SED with such an infrared excess, based on the starburst galaxy Mrk33. Our SED is based on the observed mid-IR spectrum\footnote{available for download at http://sings.stsci.edu/} of Mrk33 from SINGS observations with IRS as well as broadband photometry from Dale et al.\ (2005) to constrain the dust blackbody at longer wavelength. This SED and the broadband data appear in Fig.\ \ref{fig:modelseds}. We substituted this SED for a comparable fraction (16\%) of the dusty late-type galaxies in our model by replacing the Scd galaxies (Table \ref{t-lf}) with this starburst population. This component is shown as the dash-dot line in Fig.\ \ref{fig:24cnts}.


We find that including these sources in our PLE model yields a significant improvement in the fit to the counts, particularly at the bright end, suggesting that at 24\micron\ the counts are dominated by sources with warmer dust and higher SFRs -- primarily starburst galaxies.

While the bright end of the number counts are extremely well matched, at $S_{24 \rm{\umu m}}<0.5$ mJy the counts are still underpredicted by a factor of a few. This suggests that our model is deficient in the number of predicted 24\micron\ sources at high redshift, and comparing the predicted redshift distribution of the model to the observed $n(z)$ (Perez-Gonzalez et al., 2005) confirms that this is the case. 

Babbedge et al.\ (2006) suggest that while the galaxy luminosity functions at IRAC wavelengths show relatively little evolution, there is strong evolution in the 24\micron\ LF, particularly out to $z\approx1$. 
Including our Mrk33 component brings our model in line with local observations and in doing so enables us to match the bright-flux, low-redshift end of the counts. Allowing this warm-dust, high-SFR component to evolve more strongly could make it possible to accurately fit the 24\micron\ counts across the full range of flux; this would be in line with the approach taken in previous analyses such as that of Babbedge et al.\ (2006) and Rowan-Robinson (2001; 2009). However, we explore here the alternative possibility that the counts can be matched by exploiting the negative 24\micron\ $k$-correction, combined with the moderate PLE which was successful at IRAC wavelengths, instead of invoking much stronger evolution.

This is motivated initially by the finding that the shape of the predicted source counts for the normal spirals matches well the shape of the data, showing an upturn around 0.5--1.0 mJy where the counts briefly increase at a super-Euclidean rate (Fig.\ \ref{fig:24cnts}). The Mrk33 model, although matching the bright end normalisation well, has a much flatter shape with no increase at the faint end.

The faint-end increase in the spiral model arises from a strong negative $k$-correction, whereas in the Mrk33 SED the $k$-correction is positive. As the SEDs are redshifted, the PAH region moves into the 24\micron\ band. As this happens, the cool-dust spiral model becomes brighter (giving a negative $k$-correction) but the Mrk33 model, which has a warm dust component, becomes fainter overall. (See Fig.\ \ref{fig:modelseds}).

Therefore, where we previously used the Mrk33 component, we now use instead a population with the same $z=0$ LF as the Mrk33 population (i.e. the same $K-[24]$ colour), providing the good fit to the bright end counts, but with the $k+e$ evolution of our normal spiral galaxies. We find this gives a good fit to the data across the full flux range (Fig.\ \ref{fig:24cnts}). We are essentially using a hypothetical SED which has the same shape as the rest of our spiral galaxies but a redder $K-[24]$ colour and a greater total IR luminosity.

Using the appropriate bright end normalisation, then, PLE yields a good fit across 3 decades of 24\micron\ flux. Since there is a decoupling of the stellar and dust emissions, this arbitrary normalisation is not unphysical, however unlike our initial inclusion of the Mrk33 component it is not observationally motivated. 

We conclude that the 24\micron\ band is the first regime where our simple bimodal PLE model begins to break down. Even with the Mrk33 galaxies included, the model does not have the sufficient ingredients to match the faint-flux, high-redshift counts. However, we have shown that these data can be well matched if one component of our spiral galaxies is altered to be much redder in $K-[24]$, although the data can alternatively be matched by allowing for much stronger evolution, above the level found in our PLE model.


\begin{figure*}
\includegraphics[width=150.mm]{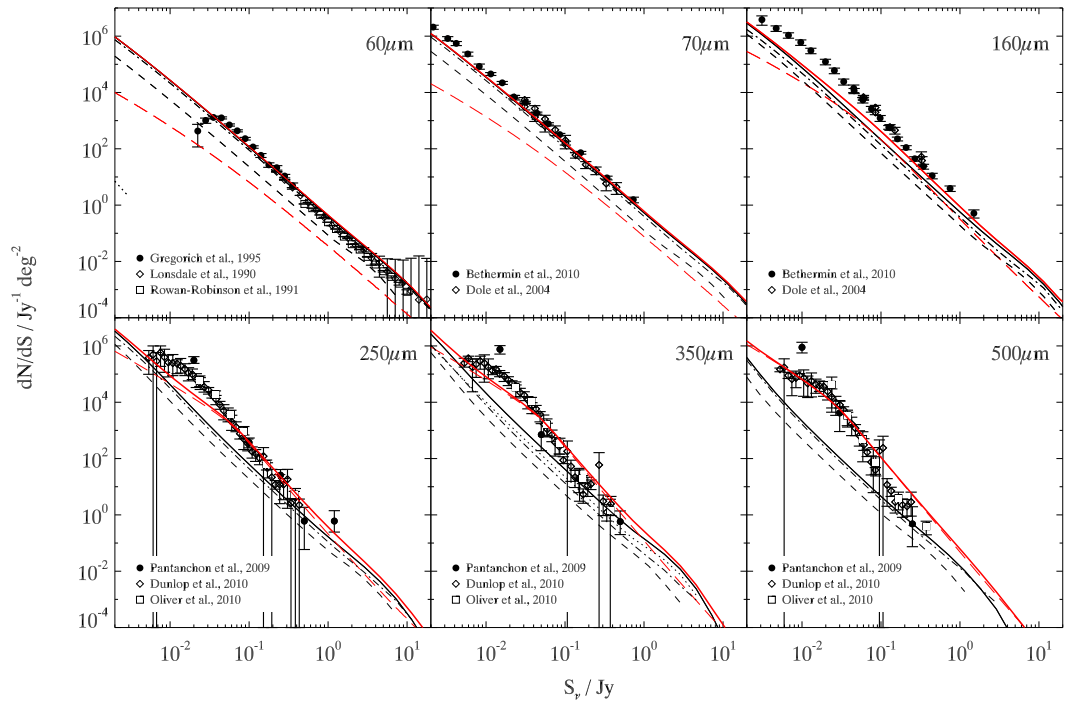}
\caption{Differential number counts at 60--500 microns. We show the prediction of our galaxy model (solid black line) as well as its constituent components -- normal spirals (dashed black) and the Mrk33 component (dot-dashed black). This model gives a reasonable fit to the bright-end data (except at 160\microns) but underpredicts the faint end by an increasingly large factor. We show also the predicted contribution from an obscured AGN model (long-dashed red) and the combined galaxy+AGN prediction (solid red), which dramatically improves the fit at long wavelengths.}
\label{fig:fircnts}
\end{figure*}


\begin{figure*}
\includegraphics[width=150.mm]{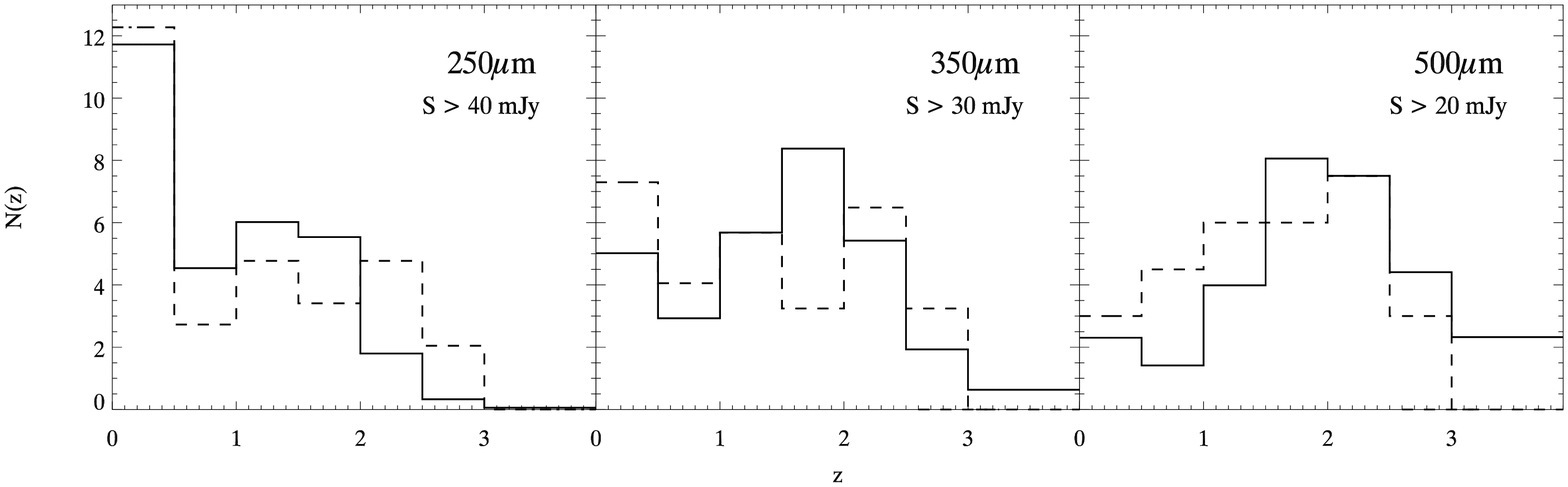}
\caption{Redshift distributions at 250\microns, 350\microns\ and 500\microns, showing our model prediction (for galaxies and obscured AGN combined; solid line) against the observed distributions published by Chapin et al. (2010; dashed line). The histograms have been normalised to the same total number of sources. In all three bands there is good agreement between the model and the data.}
\label{fig:submmnz}
\end{figure*}


\section{Far-IR/sub-mm number counts} \label{sec:countsfir}

In Fig.\ \ref{fig:fircnts} we show number counts in 6 wavebands in the far-IR and sub-mm regimes. Counts at 60\microns\ are from wide-field IRAS surveys (Lonsdale et al., 1990; Rowan-Robinson et al., 1991; Gregorich et al., 1995), 70\micron\ and 160\micron\ data come from more recent deep \emph{Spitzer} surveys (Dole et al., 2004; B\'ethermin et al., 2010) and counts at 250--500\microns come from BLAST (Patanchon et al., 2009) and \emph{Herschel} (Oliver et al., 2010; Dunlop et al., 2010) observations.

\subsection{Galaxy model predictions}

In the far-IR (upper three panels) there appears to be a continuation of the result we found at 24\microns: the galaxy model (solid black line), which continues to be dominated by the Mrk33 starburst component (dot-dashed black line), matches the bright end of the counts well, but perhaps begins to underpredict the faint end -- this can be seen in the faintest 70\micron\ counts. Our result at 60\microns\ is consistent with that of Busswell \& Shanks (2001), who also achieved a good fit to the 60\micron\ counts using a similar model.

While the fit to the 60--70\micron\ counts is, overall, reasonable, the model does very poorly at matching 160\micron\ counts, with even the bright end underpredicted. Moving to submillimetre wavelengths (lower three panels), although the  bright end appears to match well, at faint fluxes the model underpredicts the data by as much as 1--2 orders of magnitude. Therefore, normal galaxies which dominate observations at optical to mid-infrared wavelengths account for only a few percent of sources seen in the sub-mm.

\subsection{Obscured AGN model predictions}

The conventional view of the submillimetre-bright galaxy population observed at 850\microns\ is that they are high-$z$ ULIRGs with very high levels of star-formation fuelling their extreme bolometric luminosities (Smail et al., 1997; Barger et al, 1998; Alexander et al., 2005). However the possibility remains that obscured AGN make a major contribution to this population (Almaini et al., 1999; Gunn \& Shanks, 1999): in a recent paper (Hill \& Shanks, 2010) we demonstrated that a model of obscured AGN could  give a good fit to the observed 850\micron\ source counts and extragalactic background. We showed that the predictions of this model for the sub-mm contribution made by X-ray-detected AGN were in line with our measurements for X-ray sources in the Extended \emph{Chandra} Deep Field South.

The model we use is that of Gunn \& Shanks (1999), consisting of a obscured AGN enshrouded with a flat distribution of column densities between $N_{\rm{H}}=10^{19.5}$ cm$^{-2}$  and $10^{25.5}$ cm$^{-2}$ and dust masses inferred from a Galactic gas to dust ratio. This non-unified AGN model is known to match measurements of the X-ray background (Gunn, 1999). As with our galaxy model we assume a dust temperature of $T_d=30$K for the AGN and all radiation absorbed by the obscuring medium is assumed to be isotropically re-radiated in the infrared. The population evolves through PLE only.

This model gives a good fit to 850\micron\ number counts (Hill \& Shanks, 2010). Although it underpredicts the faint end, at this point our model of normal galaxies makes up the counts well. We now apply the obscured AGN model to shorter wavelengths, keeping all parameters the same as in our previous paper. In Fig.\ \ref{fig:fircnts}, the contribution from the AGN is shown as a red long-dashed line, and the combined AGN and galaxy model is shown as a solid red line. At 60\micron\ and 70\micron, the AGN are entirely negligible across the full flux range. At 160\microns, their contribution is roughly equal to that of the galaxies, however the combined galaxy-AGN model still falls short of the number counts.

At 250--500\microns, we find that the AGN model can dramatically improve the fit to the data. At 250\micron\ and 350\micron, the prediction at faint fluxes ($S<30$ mJy) is still lower than the data, but sources here are beyond the confusion limit (Nguyen et al., 2010) and so the data must be treated somewhat cautiously. At 500\microns, the bright end is a little overpredicted by the AGN model, however we suggest that overall the model matches the sub-mm data with reasonable success, and we emphasize that no aspect of the model has been altered to fit these source counts.

In Fig.\ \ref{fig:submmnz} we show predicted sub-mm redshift distributions compared to data from a BLAST survey (Chapin et al., 2010). The agreement between the model and the data is generally very good. Dunlop et al.\ (2009) show that the 250\micron\ $N(z)$ is bimodal, with the population divided broadly into low-$z$ spirals and high-$z$ submillimetre galaxies. This is replicated in our model -- the low-redshift peak arises from our spiral galaxies and the $z>0.5$ sources are made up of the obscured AGN.

In summary, our PLE galaxy model continues to perform reasonably well out to $\lambda \approx 70$\microns, but is unable to match source counts at longer wavelengths. At this point we invoke the PLE model of obscured AGN which we have previously shown to match 850\micron\ data, and find that it also gives a reasonable fit at 250--500\microns.


\section{Discussion} \label{sec:disc}

We have shown that our basic PLE galaxy model can give a good description of
galaxy populations out to 8\microns, with starlight dominating the counts out to
$\approx6$\microns\ and dust emission dominating after that out to 8 microns.
With a starburst component included in line with local observations, the model
can continue to match observations out to $\approx70$\microns\ with reasonable
success. A caveat is that the 24--70 micron counts have been fitted here with a
K-correction that comes from a hybrid galaxy template, formed from a stellar blackbody
and a PAH component in a ratio not found in local galaxies; others have
interpreted data at these wavelengths  as requiring strong evolution and this remains a
viable alternative interpretation. Either way, the success of the PLE model in 
fitting data at these long wavelengths is still impressive, given that PLE 
represents the simplest approximation other than no-evolution models.

There is an increasing consensus surrounding the hierarchical galaxy evolution
scenario, in which galaxies form at late times through a gradual build-up of
stellar mass via the merging of smaller systems. This motivates the question as
to whether PLE models should be treated simply as a convenient phenomenological
description of galaxy populations, which can match -- and, more valuably,
predict -- observations, or whether they may in fact have a real, physical
basis. In the latter case, the implication would be that galaxies form at
high-$z$ in a monolithic collapse event, and passively evolve thereafter. But if
these models represent a phenomenological description only, then it must be that
hierarchical galaxy evolution is able to present the appearance of generally
passive evolution. 

The detection of massive galaxy populations at $z>2$ has proven a challenge for
hierarchical models (Baugh et al., 1998; Somerville, Primack \& Faber, 2001; Bower 
et al., 2006) since hierarchical formation should intuitively produce few massive structures 
at high-redshift given that the process of formation is one of gradual build-up. 
However, it has been shown that massive galaxies at $z\ga2$ can be explained 
by invoking AGN feedback (e.g.\ Bower et al., 2006)
and/or cold gas accretion (Dekel et al., 2009; Kang et al., 2010). At lower redshift, 
$z<1$, hierarchical models predict a rapid evolution in the
number of massive systems, with perhaps a tenfold drop in the number of massive 
($M>10^{11}M_{\odot}$) galaxies between $z=1$ and the present (Baugh et al., 2003). 
Observations, on the other hand, have shown no evidence for density evolution of massive 
$\sim L^{*}$ ellipticals out to $z=0.9$  (Wake et al., 2006; Brown et al., 2007; Cool et al., 2008). 
However, it has been possible to reconcile the slow evolution of  observed early-type galaxies
with the hierarchical model by use of a suitable Halo Occupation Distribution (HOD).

PLE models are entirely consistent with the presence of massive galaxies at high-$z$ and
the small amount of evolution $z<1$, however these models have had only
mixed success in matching the clustering of massive galaxies. 
Wake et al.\ (2008) suggest that the evolution in the clustering of luminous red galaxies 
(LRGs) between $z=0.6$ and $z=0.2$ rules out passive evolution, particularly at
small scales, finding that a hierarchical model with a low merger rate provides
a better fit. However, Sawangwit et al.\ (2009) show that a passive model, with
no change in comoving source density, is fully compatible with their clustering
measurements in a much larger LRG sample. Thus PLE has not only been found to be an
excellent phenomenological fit to the data, it is still possible that it may even 
represent a good physical model as well.

At  wavelengths of 160\microns\ and longer our galaxy PLE model finally fails. We
then invoked the non-unified, obscured AGN model of Gunn \& Shanks (1999) which fits
the 870 micron source counts at bright fluxes, leaving the galaxy model to
contribute at fainter fluxes (Busswell \& Shanks, 2001; Hill \& Shanks, 2010). 
We showed that this model also fits the \emph{Herschel} counts at 250, 350 and 500\microns. 
Given that this could remove the need for high luminosity, massive starbursts to fit the sub-mm
counts, then this AGN model could be said to ease the problem that such massive
galaxies at high redshifts represent for the standard $\Lambda$CDM cosmological
model. There would be for example no need to invoke a top heavy IMF to explain
such sources (Baugh et al., 2005). It would only remain to check whether the model
could accommodate the galaxy PLE phenomenology at shorter wavelengths
($\lambda<160$\microns).

The AGN model we use is non-unified, with absorbed AGN having brighter sub-mm fluxes
than unabsorbed AGN (see also Page et al., 2004). Hill \& Shanks (2010) 
suggested that if a unified rather than non-unified AGN model was applied, the
AGN contribution to the sub-mm would drop dramatically
and another component would be required to fit the sub-mm source counts.
Conventionally, this is taken to be a population of highly evolved  starburst
galaxies with the bolometric luminosities of QSOs. We now consider this
alternative possibility in the context of a PLE model where these sources are
modelled as dust-obscured early-type progenitors. We have therefore
modelled a population of early-types that evolve according to a
$\tau=1$ Gyr model with a Chabrier IMF and dust absorption in the range
$A_B=0.3$--1 mag, which re-radiates optical light into the FIR assuming the usual
30K dust temperature. Although this model produces a K band evolution of 2--3 mag
at $z\approx2$--6 relative to the $x=3$ IMF model or the $k$-correction, we found the
contribution to the sub-mm counts to be negligible.

In addition, the Chabrier IMF produces strong evolution at $z\approx1$ and would need
a dust absorption of $A_B\approx2$ mag to avoid over-predicting the high-$z$ tail
of the $K<20$ redshift distribution. This level of extinction would destroy the
viability of the PLE models to fit galaxy colour and count distributions through
the optical and NIR bands. Appealing to a top-heavy IMF might still enable
the sub-mm counts to be fitted, but such an IMF could not be allowed to apply at
$z\approx1$--2 because the problem of overpredicting the K20 $n(z)$
would be exacerbated. Somewhat paradoxically, although the semi-analytic model 
that invokes a top-heavy IMF (see Baugh et al., 2005; Lacey et al., 2008) 
fits well the sub-mm counts, it appears to seriously \emph{underpredict} 
the average K20 redshift (Gonzalez-Perez et al., 2009). These authors also show 
that the model of Bower et al.\ appears to fit the $K$ $n(z)$ better, but fails to fit 
the sub-mm counts.

Certainly in the context of our PLE model there is greater observational motivation
for a dwarf-dominated IMF than a top-heavy IMF: the top-heavy IMF is not required since
sub-mm counts are matched by the AGN population, and the dwarf-dominated IMF can therefore 
provide an explanation for the absence of a detectable initial burst of
star-formation from early-type galaxies at very high redshift -- this burst may be absent due to a lack
of bright stars in the IMF of early-types. Arguments against such a model
include the apparently prompt enrichment of early type galaxies evidenced by
their $\alpha$-element enhancement (Tinsley, 1979; Trager et al., 2000). But the excellence of the
fits of our simple galaxy count models from the UV through the K band to the
sub-mm means that this alternative idea for the origin of early-types is still worthy
of serious consideration.


\section{Summary} \label{sec:summ}

In this work we have explored how successfully the phenomenology of galaxy
populations at mid-IR, far-IR and sub-mm wavelengths can be described by a PLE
galaxy model known to fit observations in the optical/NIR. We employed the
simple bimodal model developed by Metcalfe et al.\ (1995, 2001, 2006; see also
McCracken et al., 2006), which includes spiral galaxies with a very modest
amount of dust ($A_B=0.3$ mag) and early-type galaxies with no dust absorption.
Our conclusions are as follows:

\begin{itemize}

\item This simple model continues to provide a reasonably good description of galaxy
populations out to 8\microns. At 6--8\microns\ the bright source counts move
from being starlight dominated to dust emission dominated. At 8\microns, stellar
light makes only a $\approx10$\% contribution at bright fluxes ($>400$ $\umu$Jy) but a
$\approx90$\% contribution at very faint fluxes ($<4$ $\umu$Jy).

\item Colour-redshift tracks from the model are well matched to galaxy colours
out to at least $\lambda=8$\microns\ and $z=2.5$.

\item At 24--70\microns, our basic model no longer matches observed source
counts, accounting for only $\approx20$\% of 24\micron\ sources. By replacing
16\% of our spirals with warmer-dust, higher-SFR galaxies, consistent with local
observations, we match the bright counts very well, but there remains an
underprediction of the faint-end counts. If we instead use a population of
spirals which have our standard late-type SED but the $z=0$ $K-[24]$ colour of
Mrk33 (essentially, the normal spiral SED normalised to brighter flux), we fit
the counts across the full flux range.

\item At sub-mm wavelengths ($\lambda=250$, 350, 500\microns) the galaxy PLE model
fails -- we find that normal galaxies account for only a few percent of sources
here. However, we show that a PLE model of obscured AGN, which we have
previously shown to give a good fit to observations at 850\microns\ (Hill \&
Shanks, 2010), can again give a good fit to the counts and redshift
distributions in these bands.

\end{itemize}


\section*{Acknowledgements}

MDH gratefully acknowledges the receipt of an STFC studentship.

\label{lastpage}

\end{document}